\documentstyle[preprint,eqsecnum,aps,prl]{revtex}
\newcommand{\cro}{Cd$_2$Re$_2$O$_7$}
\newcommand{\msr}{$\mu$SR}

\begin{document}

\title
{Quasiparticle Excitation in the Superconducting Pyrochlore
 \cro\ Probed by Muon Spin Rotation
}

\author
{ 
Ryosuke {\sc Kadono}\footnote{Also at School of Mathematical and Physical Science, 
The Graduate University for Advanced Studies}, Wataru {\sc Higemoto},
Akihiro {\sc Koda}, Yu {\sc Kawasaki}$^1$, 
Masashi {\sc Hanawa}$^2$, and Zenji {\sc Hiroi}$^{2}$
}

\address
{
Institute of Materials Structure Science, High Energy Accelerator Research Organization (KEK),
Tsukuba, Ibaraki 305-0801\\
$^1$Graduate School of Engineering Science, Osaka University, Toyonaka, Osaka 560-8531\\
$^2$Institute for Solid State Physics, University of Tokyo, Kashiwa, Chiba 277-8581
}

\date
{Received September 12, 2001
%\today
}

\maketitle

\begin{abstract}
%{
The quasiparticle excitations in the mixed state of 
\cro\ have been studied by means of muon spin rotation/relaxation (\msr).
The temperature dependence of the magnetic penetration
depth ($\lambda$) is consistent with a nearly isotropic
superconducting order parameter, although a slight discrepancy 
which is dependent on the details in the analysis may be present.
This is also supported by the relatively weak field dependence of $\lambda$.
%}
%\kword
%{
%superconductivity, penetration depth, pyrochlore, \msr
%}
\end{abstract}

%\begin{document}
%\sloppy
%\maketitle

\newpage

A class of metal oxides isostructural to mineral pyrochlore 
has been attracting considerable attention because they exhibit a wide variety of
interesting physical properties.\cite{Subramanian:83} 
The pyrochlore has a general formula of A$_2$B$_2$O$_7$,
consisting of BO$_6$ octahedra and
eightfold coordinated A cations, where A and B are
transition metals and/or rare-earth elements.  In particular,
the B sublattice can be viewed as a three-dimensional network
of corner-sharing tetrahedra, providing a testing ground for studying
the role of geometrical frustration in systems which have local spins at 
B sites with antiferromagnetic (AFM) correlation.\cite{Ramirez:94}
Such systems are known to remain frustrated even when the
exchange interaction is ferromagnetic (FM), provided that
the spin correlation has local Ising anisotropy.\cite{Harris:97}
Recent studies have revealed a rich variety of phenomena seemingly related to 
the geometrical frustration, such as the occurrence of a spin-glass (SG) phase
in $R_2$Mo$_2$O$_7$ with $R$=Y, Tb and Dy\cite{Dunsiger:96,Gardner:99},
the unusual behavior of ordinary and anomalous Hall coefficients in the same compound
with $R$=Nd, Sm and Gd\cite{Taguchi:99,Katsufuji:00}, and
the ``spin-ice" phase in $R_2$Ti$_2$O$_7$ with $R$=Dy and Ho.\cite{Harris:97,Ramirez:99}

Although metallic pyrochlores comprise a minority subgroup of the pyrochlore family,
they consist of distinct members such as Tl$_2$Mn$_2$O$_7$,
which exhibits colossal magnetoresistance.\cite{Shimakawa:96,Shimakawa:99}
Moreover, the recently revealed superconductivity in \cro, a 5$d$
transition metal pyrochlore\cite{Hanawa:01,Jin:01}, demonstrates that the pyrochlores
provide a fertile field for electronic correlation adjacent to the perovskite compounds.
In this context, it is noteworthy that LiV$_2$O$_4$, a cubic spinel compound
in which the V sublattice is isostructural to the B sublattice in pyrochlore, 
behaves similarly to a heavy fermion metal.\cite{Kondo:97}

It is reported that \cro\ falls into the bulk superconducting state below $T_c\simeq 1\sim 2$ K, as 
confirmed by a large jump of specific heat $\Delta C_c$ as well as large diamagnetism 
due to the Meissner effect associated with the occurrence 
of zero resistivity.\cite{Hanawa:01}  The dc magnetization curve indicates
that the superconductivity is of type II with the upper critical field close to
0.29 T at 0 K.\cite{Hiroi:01}  The ratio $\Delta C_c/\gamma T_c$
(with $\gamma$ being the Sommerfeld constant) is reported to be 1.15,
which is smaller than the predicted value of 1.43 for isotropic BCS superconductors. 
Unfortunately, these measurements were
performed only above 0.4 K and are thus inconclusive in determining the 
detailed characteristics of superconductivity in \cro.  In this Letter, we
report on the quasiparticle excitations in the mixed state of \cro\ studied by muon spin 
rotation/relaxation ($\mu$SR).  The magnetic penetration depth $\lambda$,
which reflects the population of normal electrons (``quasiparticles") in the
superconductive state, is determined microscopically by measuring the muon
spin relaxation due to the spatial inhomogeneity of magnetic induction in the flux 
line lattice (FLL).  We show that the 
temperature dependence of $\lambda$ is more or less consistent with the prediction 
based on the BCS superconductors with an isotropic gap, although 
a slight discrepancy is suggested by the detailed analysis.

The \msr\ experiments on both single-crystal and polycrystalline \cro\ were performed on 
the M15 beamline at the TRIUMF muon facility which provides a beam of nearly 
100\% spin-polarized positive muons of momentum 28.6 MeV/c. 
The specimen was mounted on the coldfinger of a $^3$He-$^4$He dilution refrigerator 
and cooled from a temperature above $T_c$ after setting the 
magnetic field at every field point (i.e., field-cooling) to eliminate the effect of 
flux pinning.  The field and temperature scan data were obtained at $T=0.2$ K and
at $H=0.1$ T, respectively.
Muons were implanted into the specimen (measuring about 10 mm$\times$10 mm 
and 1 mm thick) after being passed through a beam collimator.
The initial muon spin polarization was perpendicular
to the magnetic field $H$ and thus 
to the FLL in the superconducting state.

Since the muons stop randomly along the length scale of the 
FLL, the muon spin precession signal $\hat{P}(t)$ provides 
a random sampling of the internal field distribution $B({\bf r})$,
\begin{eqnarray}
\hat{P}(t) &\equiv &P_x(t)+iP_y(t)=\int_{-\infty}^\infty n(B)\exp(i\gamma_\mu Bt)dB,\\
n(B)&=&\frac{d{\bf r}}{dB},
\end{eqnarray}
where $\gamma_\mu$ is the muon gyromagnetic ratio (=$2\pi\times$135.53 MHz/T),
and $n(B)$ is the spectral density for muon precession determined by the 
local field distribution.
These equations indicate that the real amplitude of the Fourier-transformed muon
precession signal corresponds to the local field distribution $n(B)$.
The local field distribution can be approximated as the sum of
magnetic induction from isolated vortices in the London model to yield
\begin{equation}
B({\bf r})=B_0\sum_{\bf K}\frac{e^{-i{\bf K}\cdot{\bf r}}e^{-K^2\xi_v^2}}
{1+K^2\lambda^2+O(K_x^2,K_y^2)}\:,
\end{equation}\label{Br}
where ${\bf K}$ is a translation of the vortex reciprocal lattice,
$B_0$ ($\simeq H$) is the average internal field, $\lambda$ is the
London penetration depth, and $\xi_v$ is the cutoff parameter.
The term $O(K_x^2,K_y^2)$ denotes the nonlocal effect in which
the electromagnetic response kernel $Q({\bf K})$ generating
the supercurrent around the vortex depends on ${\bf K}$.
While this term is eliminated in the conventional BCS
superconductors with isotropic $s$-wave pairing,
it becomes important for the moe complex order parameters 
such as anisotropic $s$-wave or $d$-wave (e.g., $d_{x^2-y^2}$).
The London penetration depth in the FLL state is
related to the second moment 
$\langle \Delta B^2\rangle=\langle (B({\bf r})-B_0)^2\rangle$ 
of the field distribution reflected in the \msr\ line shape
(where $\langle\:\rangle$ means the spatial average).  
In polycrystalline samples, 
a Gaussian distribution of local fields is a good approximation, where 
\begin{eqnarray}
\hat{P}(t) &\simeq&\exp(-\sigma^2t^2/2)\exp(i\gamma_\mu Ht),\\
\sigma&=&\gamma_\mu\sqrt{\langle\Delta B^2\rangle }.
\end{eqnarray}
For the case  of ideal triangular FLL with isotropic effective carrier mass $m^*$
and a cutoff $K\approx1.4/\xi_v$ 
provided by the numerical solution of the Ginsburg-Landau theory, 
the London penetration depth $\lambda$ (with $O(K_x^2,K_y^2)=0$)
can be deduced from $\sigma$ using the following 
relation\cite{Pincus:64,Aeppli:87,Brandt:88}, 
\begin{equation}
\sigma\ [\mu{\rm s^{-1}}] = 4.83\times 10^4(1-h) \lambda^{-2}\ [{\rm nm}],
\label{eq2}
\end{equation}
where $h=H/H_{c2}$.
While the above form is valid for $h<0.25$ or $h>0.7$,
a more useful approximation valid for an arbitrary field is \cite{Brandt:88}
\begin{equation}
\sigma\ [\mu{\rm s^{-1}}] = 4.83\times 10^4 
(1-h)[1+3.9(1-h)^2]^{1/2} \lambda^{-2}\ [{\rm nm}].
\label{eq3}
\end{equation}
In both cases, $\lambda$ is related to the superconducting carrier density
$n_s$ as 
\begin{equation}
\lambda^2=\frac{m^*c^2}{4\pi n_se^2},\label{lmd-ns}
\end{equation}
indicating that $\lambda$ is enhanced upon the reduction of $n_s$
due to the quasiparticle excitations.  For simplicity, we adopt
eq.~(\ref{eq2}) for the following analysis.

In a preliminary analysis, we found that the spin relaxation rate
due to the FLL formation is less than 0.1 $\mu$s$^{-1}$ which
is typically of the same order of magnitude as that due to static random local fields
from nuclear magnetic moments.
This means that the additional relaxation due to $^{111,113}$Cd and $^{185,187}$Re 
{\it nuclear} moments must be considered for the proper
estimation of $\lambda$ in \cro.
To this end, the following equation was
used in the actual fitting analysis of the time spectra,
\begin{equation}
\hat{P}(t) =\exp[-(\sigma_0t)^{\nu}+\sigma^2t^2/2)]\exp(i\gamma_\mu Ht),\\
\end{equation}
where $\sigma_0$ is the relaxation rate due to the nuclear moments 
and $\nu$ is the power of relaxation.
The parameters $\sigma_0$ and $\nu$ were evaluated by fitting the 
time spectra above $T_c$ with $\sigma$ set to zero, yielding 
$\sigma_0\simeq0.057(1)$ $\mu$s$^{-1}$ and $\nu\simeq 1.0$
in the polycrystalline specimen at $H=0.1$ T.  
Then, $\sigma$ due to the formation of FLL below $T_c$
was deduced by analyzing data with $\sigma_0$ and $\nu$ 
being fixed to the above values.  A similar analysis was performed
for the data obtained for the single crystals.

Figure \ref{tdep}(a) shows the temperature dependence of $\sigma$
in \cro\ at $H=0.1$ T. Upon the onset of FLL formation,
$\sigma$ exhibits a gradual increase with decreasing
temperature just below $T_c(0.1\:{\rm T})\sim0.7$ K at this field.
According to the empirical two-fluid
model approximately valid for conventional BCS superconductors,
we have
\begin{equation}
\lambda(t)=\lambda(0)\frac{1}{\sqrt{1-t^4}},
\end{equation}
which leads to
\begin{equation}
\sigma(t)=\sigma(0)(1-t^4),
\end{equation}
where $t\equiv T/T_c(0.1\:{\rm T})$.  
The fitting analysis by the same formula with an arbitrary power, 
\begin{equation}
\sigma(t)=\sigma(0)(1-t^\beta),
\end{equation}\label{sgm-t}
with $T_c$  as a free parameter 
yields $\beta=2.8(5)$ and $T_c=0.65(3)$ K. 
We also found that $T_c=0.62(1)$ K when $\beta=4$ is assumed,
which is slightly lower than the value $T_c\simeq0.7$ K estimated  
from the specific heat measurement for this field.
Although the difference is not obvious between 
these cases (solid curve for $\beta=2.8$ and
the dashed curve for $\beta=4$) in Fig.~1(a), we note that
the reduced $\chi^2$ for the former is almost two times smaller 
(better) than the latter. Taking $H_{c2}=0.29$ T, the penetration depth 
extrapolated to $T=0$ ($\lambda(0,0.1\:{\rm T}))$ is 700(8) nm,
leading to the Ginsburg-Landau parameter $\kappa\simeq21$
with $\xi\simeq34$ nm estimated from $H_{c2}(0)$. Thus,
it is concluded that \cro\ is a typical type II superconductor
with large $\lambda$,
which is consistent with the results of magnetization measurement.\cite{Hanawa:01}

The finding that $\beta\simeq2.8$ may indicate that the deviation 
$\Delta\lambda=\lambda(t)-\lambda(0)\propto T^\beta$ exhibits a tendency 
predicted for the case of line nodes ($d$-wave pairing) with
some disorder (i.e., dirty limit),  where 
$\Delta\lambda\propto T^2$.\cite{Hirschfeld:93}
Such a temperature dependence has actually been observed in high-$T_c$ (YBCO) 
cuprates.\cite{Bonn:94,Sonier:97c}  
Compared with the case of isotropic gap
$\Delta_{\hat{k}}=\Delta_0$,
the quasiparticle excitations are enhanced along nodes
($|\Delta_{\hat{k}}|=0$) to reduce 
average $n_s$, leading to the enhancement of $\lambda$. 
However, the result of the fitting analysis also suggests that the 
discrepancy may be attributed to experimental uncertainty
including the precise value of $T_c$.
Thus, we are led to conclude that the order parameter
in \cro\ is mostly isotropic with a possibility of residual
weak anisotropy as suggested by the slightly reduced value of $\beta$.

As shown in Fig.~2(a), $\sigma$ decreases with increasing external
field, where the general field dependence is determined by the
increasing contribution of normal vortex cores and the stronger overlap of
field distribution around the cores which are described by the term $(1-h)$ in
eq.(\ref{eq2}).  The fitting
analysis by eq.(\ref{eq2}) with $H_{c2}$ and $\lambda(H=0)$
as free parameters yields  $H_c2=0.37(5)$ T and 
$\lambda(H=0)=796(12)$ nm at 0.2 K.
Since the deduced value of $H_{c2}$ is consistent with that
obtained from the magnetization measurement ($\simeq0.29$ T),
we can conclude that the observed field dependence of $\lambda$ 
is mostly due to the vortex core/overlap effect, except for the fields below $\sim$0.06 T
where a slightly steeper reduction of $\sigma$ is suggested.
The penetration depth deduced for each field using eq.(\ref{eq2}) 
with $H_{c2}=0.37$ T is plotted in Fig.~2(b). 
In order to evaluate the relative strength of the pair-breaking
effect, it is useful to introduce a dimensionless parameter
$\eta$ to describe the field dependence of $\lambda$ with 
the following simple linear relation,
\begin{equation}
\lambda=\lambda({\rm 0.2 K},h)[1+\eta h], \label{lmdh}
\end{equation}
where $h\equiv H/H_{c2}({\rm 0.2 K})$ with
$H_{c2}({\rm 0.2 K})=0.37$ T.
From the analysis of data in Fig.~2(b) by eq.~(\ref{lmdh}),
we find that $\eta=0.38(14)$ with $\lambda({\rm 0.2 K},0)=741(5)$ nm
for $0\le H\le0.06$ T.

In general, the field dependence of $\lambda$ is enhanced by 
two different mechanisms, i.e., the nonlinear effect in
the semiclassical Doppler shift of the quasiparticle energy
levels due to the supercurrent around the vortex cores, and the nonlocal 
effect which further modifies the quasiparticle excitation spectrum
in the momentum space.  In particular, the nonlocal effect 
is important in the system with line nodes because
the coherence length is inversely
proportional to the order parameter, such that
$\xi_0(\hat{k})=\hbar v_F/\pi\Delta_{\hat{k}}$.
The divergence of $\xi_0$ along the nodal directions $|\Delta_{\hat{k}}|=0$
means that the response of quasiparticles near the
nodes is highly nonlocal. 

It is predicted that $\eta\ll 1$ for the isotropic $s$-wave pairing
because the finite gap prevents the shifted levels of quasiparticle excitations
from being occupied at low temperatures.  This is supported, for example, by
the recent observation in CeRu$_2$, in which $\eta\simeq0$ over
the field region $0\le h\le0.5$ where the system behaves more or less
as a conventional BCS superconductor with isotropic $s$-wave pairing.\cite{Kadono:01}
On the other hand, stronger field dependence is predicted for the 
case of $d$-wave or anisotropic $s$-wave pairing due to the
excess population of quasiparticles in the region where $|\Delta_{\hat{k}}|$
is small or zero.  Typical examples for the $d$-wave
pairing are those of high-$T_c$ cuprates
in which $\eta$ is reported to be 5$\sim$6.6 for YBCO.\cite{Sonier:97a}
Meanwhile, in the case of an $s$-wave superconductor YNi$_2$B$_2$C in which   
strong anisotropy for $\Delta_{\hat{k}}$ is suggested experimentally\cite{Yokoya:00},
$\eta\simeq1$ is reported from a \msr\ study.\cite{Ohishi:01}
The comparison of these earlier results with our result suggests 
that the anisotropy of the order parameter in \cro\ is considerably
smaller than YNi$_2$B$_2$C.   

Finally, we comment that the recent observation of 
a clear coherence peak below $T_c$ in the $^{187}$Re NQR measurement\cite{Vyaselev:01}
does not necessarily indicate the absence of anisotropy in the order parameter.  
In the case of the anisotropic energy gap,
the magnitude of the coherence peak depends on the mean free path $l$. 
The coherence peak is enhanced in a certain condition of $l$ 
where quasiparticles probe only a limited region of the Fermi surface. 
Thus, while the actual value of $l$ is difficult to estimate because of the semimetallic 
character of this compound, the NQR result does not completely rule out the
presence of anisotropy in general.

In summary, we have investigated the quasiparticle excitations in the 
mixed state of \cro\ by \msr.
The temperature and field dependence of the London penetration
depth indicates that the basic feature is consistent with the
isotropic order parameter for BCS $s$-wave pairing, although there remains a certain subtlety
suggested by the small deviation from the theoretical prediction
which may be better understood by considering a weak anisotropy.

We would like to thank the staff of TRIUMF for their technical
support during the experiment.
This work was partially supported by a
Grant-in-Aid for Scientific Research on Priority Areas and
a Grant-in-Aid for Creative Scientific Research from the Ministry of
Education, Culture, Sports, Science and Technology of Japan.

%\newpage

%\newpage
%Figure Captions

\begin{figure}
%\figureheight{11cm}
%\begin{center}
%\mbox{\epsfxsize=0.4\textwidth \epsfbox{cro-fig1.ps}}
%\end{center}
\caption{Temperature dependence of a) the muon spin relaxation
rate $\sigma$ due to flux line lattice and b) the magnetic penetration 
depth $\lambda$, where solid circles are the data from the polycrystalline
specimen and open triangles are those from single crystals. 
Solid curves are results of fitting by a relation 
$\sigma\propto1/\lambda\propto1-(T/T_c)^\beta$
with $\beta$ and $T_c$ being free parameters. Dashed curves are
obtained when $\beta=4$ with $T_c$ as a free parameter.
}\label{tdep}
\end{figure}

\begin{figure}
%%\figureheight{11cm}
%\begin{center}
%\mbox{\epsfxsize=0.4\textwidth \epsfbox{cro-fig2.ps}}
%\end{center}
\caption{Magnetic field dependence of a) the muon spin relaxation
rate $\sigma$ due to flux line lattice and b) the magnetic penetration 
depth $\lambda$. The solid line in a) is a result of fitting by eq.(\ref{eq2})
with $H_{c2}$ being a free parameter. Dashed lines are obtained
by fitting data with a relation $\lambda\propto1+\eta(H/H_{c2})$ only for
$0\le H\le 0.06$ T.
}\label{fdep}
\end{figure}

%
%\newpage
%\begin{center}
%\mbox{\epsfxsize=0.85\textwidth \epsfbox{cro-fig1.ps}}\\
%\vspace{1cm}
%Fig.1\\ Ryosuke Kadono
%\end{center}
%\newpage
%\begin{center}
%\mbox{\epsfxsize=0.85\textwidth \epsfbox{cro-fig2.ps}}\\
%\vspace{1cm}
%Fig.2\\ Ryosuke Kadono
%\end{center}

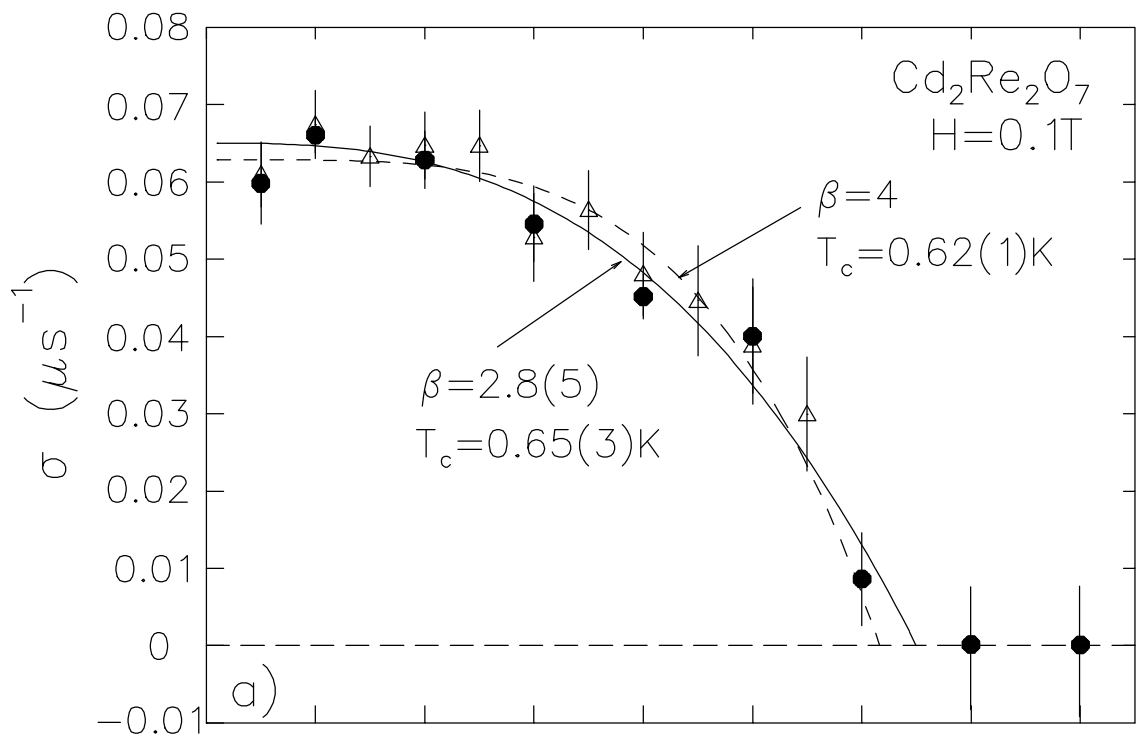
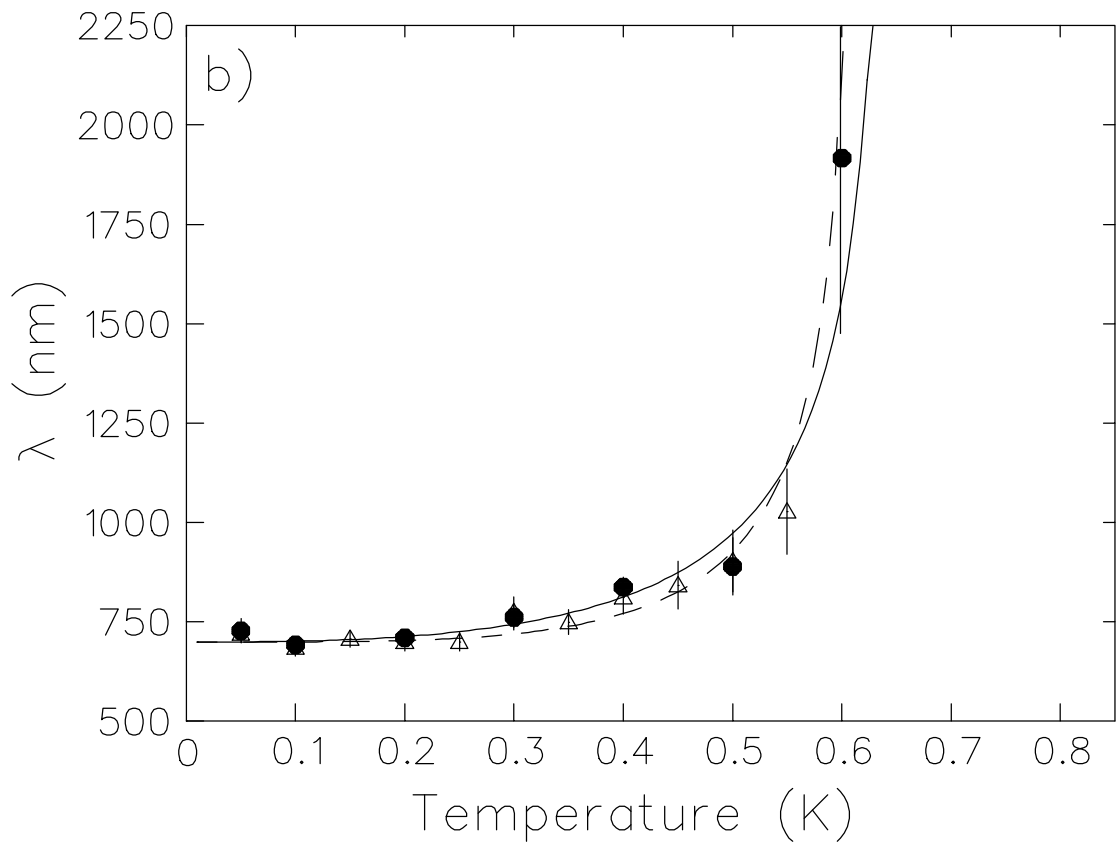
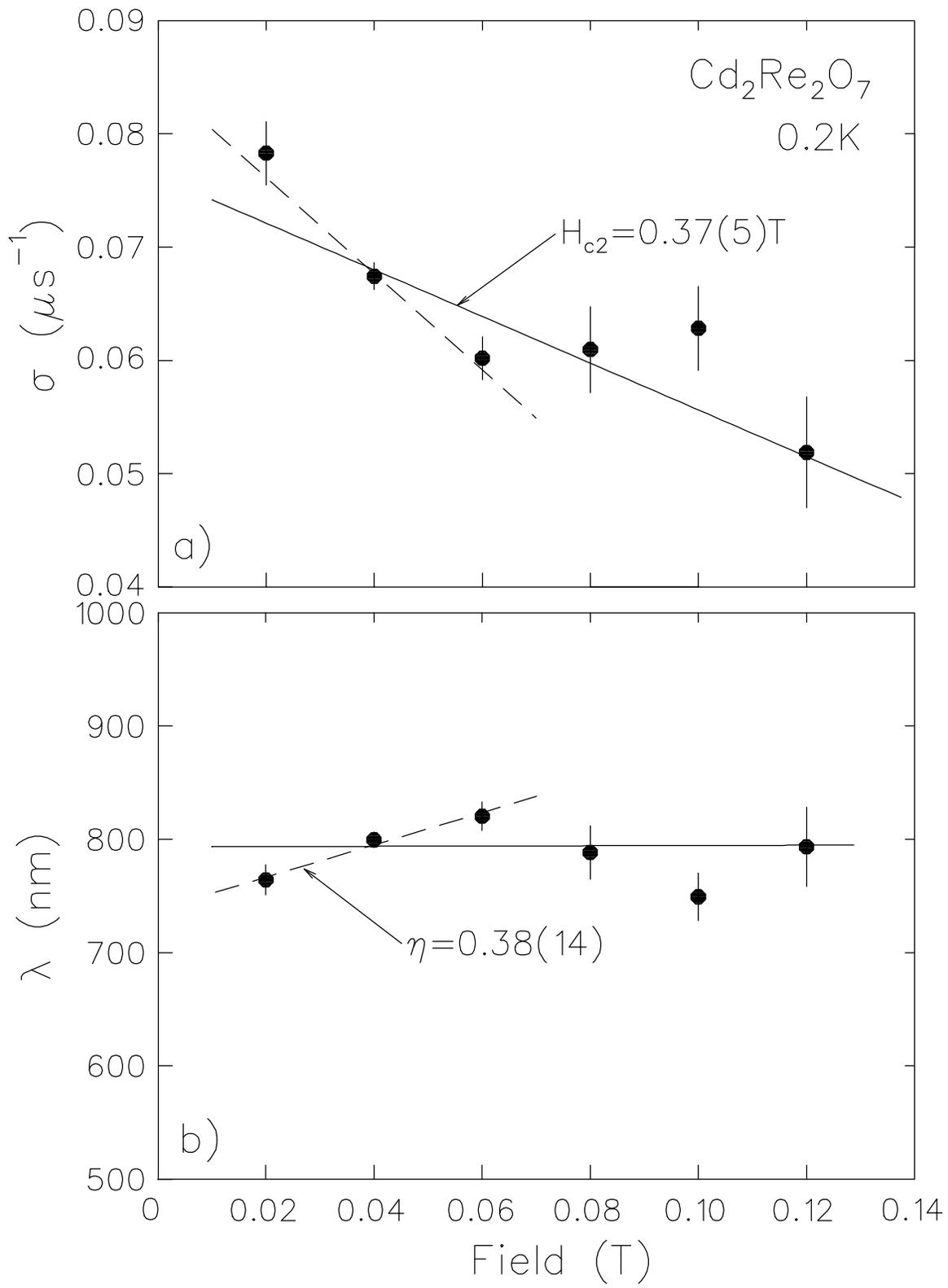
\end{document}